\documentclass[9pt,twocolumn,twoside]{opticajnl}
\journal{opticajournal} % use for journal or Optica Open submissions

\setboolean{shortarticle}{true}
\usepackage{graphicx}
\usepackage{lineno}
\linenumbers % Turn off line numbering for Optica Open preprint submissions.

\title{Multi-NA Metalens Array for Compact High-NA Microscopy}

\author[1,*]{Alireza Khalilian}
\author[2]{Jie Fan}
\author[3]{Mehdi Sh. Yeganeh}
\author[3]{Joe Lo}
\author[1,4]{Yasha Yi}

\affil[1]{Integrated Nano Optoelectronics Laboratory, University of Michigan, 4901 Evergreen Rd., Dearborn, Michigan 48128, USA}
\affil[2]{Department of Natural Sciences, College of Arts, Sciences, and Letters, University of Michigan-Dearborn, 4901 Evergreen Rd, Dearborn, MI 48128, USA}

\affil[3]{Department of Mechanical Engineering, University of Michigan-Dearborn, Dearborn, USA}
\affil[4]{Energy Institute, University of Michigan, 2301 Bonisteel Blvd., Ann Arbor, Michigan 48109-2100, USA}

\affil[*]{akhalili@umich.edu}

\begin{abstract}
We demonstrate a CMOS-compatible silicon-rich nitride metalens array for visible microscopy at \(\lambda\approx 660\,\mathrm{nm}\). Three co-planar elements provide numerical apertures of 0.54, 0.92, and 0.97, enabling within-sample NA benchmarking without objective swaps. Imaging of annular cell monolayers labeled in the AF647 ZO-1 channel shows progressive sharpening of junctional edges and reduced blur with increasing NA, consistent with diffraction-limited scaling. The compact, planar platform supports direct integration with image sensors, offering a practical route to high-NA fluorescence bioimaging in space-constrained, low-cost systems.
\end{abstract}

\setboolean{displaycopyright}{false} % Do not include copyright or licensing information in submission.

\begin{document}
\nolinenumbers
\maketitle

\section{Introduction}
High resolution cell imaging is central to modern biology and pathology, yet conventional high numerical aperture microscope objectives are bulky, alignment sensitive, and costly to scale into compact or disposable instruments\cite{way2023evolution, pan2022dielectric}. Flat metasurface lenses enable diffraction-limited focusing and wavefront control in sub-millimeter form factors and have matured rapidly for visible-band imaging. Recent demonstrations span fluorescence excitation and collection as well as quantitative phase modalities, validating metalenses as practical building blocks for miniaturized bio-imaging systems \cite{peng2024metalens,javed2025nearly, sun2025high}. At visible wavelengths, \(\mathrm{TiO_2}\) metalenses have reached \(\mathrm{NA}\approx 0.8\) with measured focusing efficiencies on the order of \(60\text{--}85\%\), establishing performance comparable to commercial objectives in a much thinner form factor \cite{khorasaninejad2016metalenses, khorasaninejad2016polarization}. Metalens arrays have also been integrated into microscope architectures to expand field of view and improve image quality, further underscoring suitability for biological samples \cite{ye2022chip, xu2020metalens}.

Translating such optics into robust bio-imaging modules requires a platform that combines high visible-band index \(n\), low extinction \(k\), wafer-level repeatability, and compatibility with silicon-based manufacturing. \(\mathrm{TiO_2}\) offers low loss across the visible and supports high NA with mature libraries, although many implementations rely on ALD and high aspect-ratio pattern transfer that can limit throughput in standard back end of line (BEOL) \cite{munoz2019speeding}. GaN provides similarly low loss and strong focusing performance, yet typically depends on epitaxial growth and heterogeneous integration that are not uniformly available in mainstream CMOS fabs \cite{chen2021high, chen2017gan}. Amorphous silicon, including hydrogenated variants, is attractive for its very high index and low-temperature PECVD deposition. With well controlled recipes its red-band loss can be acceptable and its compact phase shifters benefit extreme NA or tight pitches. The trade space includes recipe-sensitive \(k\) near 633–660\,nm, light and heat history that can shift \(n\) and \(k\), Cl\(_2\)/HBr etch stacks that favor thicker hard masks and careful roughness control, and chamber housekeeping to suppress powder formation at high silane rates \cite{tsai1993amorphous, i2002growth, staebler1977reversible}. Silicon-rich nitride offers a complementary route that aligns closely with CMOS BEOL practice. By tuning the Si:N ratio, SRN provides moderately high, composition-adjustable \(n\) with low \(k\) in the red, deposited at low temperature by PECVD and patterned with fluorine-based ICP-RIE using thin hard masks. These attributes facilitate subwavelength pitches at moderate aspect ratio, stress control and uniformity across wafers, and single-layer fabrication of multi-NA arrays with predictable yield \cite{goldberg2024silicon, khalilian2025polarization, de2024silicon, fan2018silicon, zhang2025chip}. The principal trade is a lower \(n\) than a-Si, which modestly increases the required lateral feature size or height for a given \(2\pi\) phase, and the need to avoid overly Si-rich films that raise \(k\). In this work we select SRN because its tunable \(n\)–\(k\) window at \(\lambda\approx 660\,\mathrm{nm}\), fluorine-etch compatibility, and wafer-scale stability match the goal of co-fabricating compact, high-NA metalens arrays intended for CMOS-adjacent bio-imaging systems \cite{kanellopulos2024stress, mcclung2024visible}.

Building on this rationale, we target the red channel at \(\lambda\approx 660\,\mathrm{nm}\) to align with widely used far-red fluorophores and reduced tissue scattering \cite{berlier2003quantitative}. We design a three-element, co-planar metalens array that provides numerical apertures of 0.54, 0.92, and 0.97 using a library-driven synthesis under the locally periodic approximation, and we realize the layout in a single SRN layer suitable for wafer-level processing. We validate high-NA focusing and imaging on standard resolution targets and on annular cell monolayers labeled in the AF647 ZO-1 channel, enabling within-sample benchmarking of NA without objective swaps. The results position SRN as a CMOS-aligned, low-loss platform for compact, high-NA visible microscopy.

\section{Theory}
\begin{figure}[!t]
  \centering
  \includegraphics[width=8.7cm]{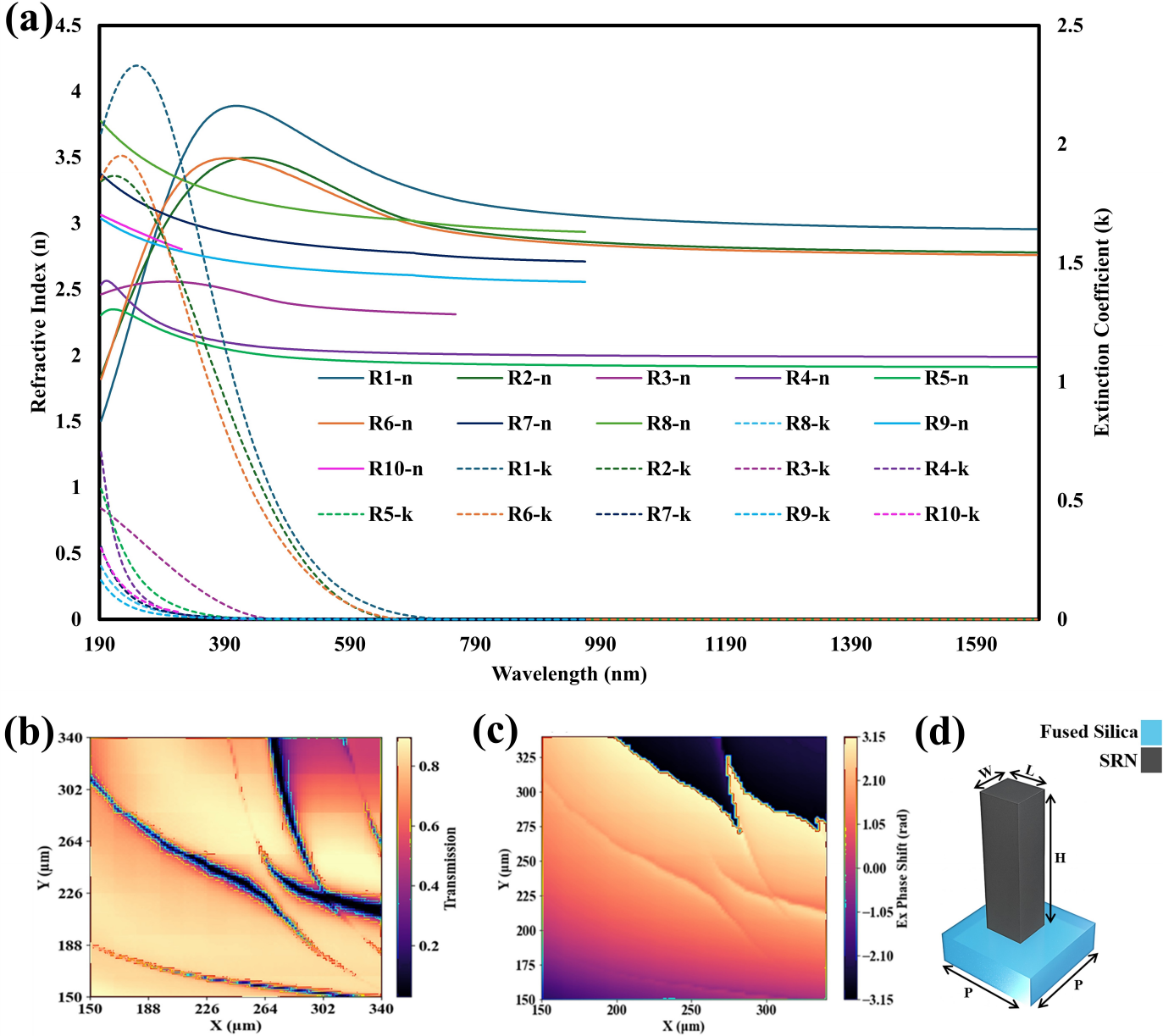}
  \caption{(a) Woollam ellipsometry showing SRN \(n\) (solid) and \(k\) (dashed) for recipes R1--R10 on fused silica. (b) Transmission map at \(660\,\mathrm{nm}\) for the chosen recipe. (c) Phase \(\phi\) at \(660\,\mathrm{nm}\) over the same design space. (d) Unit cell with SRN pillar of width \(W\), length \(L\), height \(H\), and pitch \(P\); lateral sizes \(W,L\in[150\text{--}340]\,\mathrm{nm}\).}
  \label{fig:unitcell}
\end{figure}

We design the metasurface by selecting unit cells from a precomputed library under the locally periodic approximation (LPA). The aperture is sampled on a square lattice with pitch \(P\) and a fixed height \(H\) in silicon-rich nitride (SRN). Each unit cell contains one rectangular pillar with in-plane sizes \((L,W)\). For a chosen input polarization at wavelength \(\lambda\), the library provides the complex transmission
\begin{equation}
t(L,W)=\sqrt{T(L,W)}\,e^{i\phi(L,W)}
\end{equation}
on a grid of \((L,W)\). The library is generated with unit-cell FDTD using periodic boundaries in the lateral directions and perfectly matched layers (PML) along the optical axis, with a normally incident plane wave. The simulated phase is unwrapped to remove \(2\pi\) jumps, wrapped into \([0,2\pi)\), and interpolated together with \(T\) so that \(\tilde{\phi}(L,W)\) and \(\tilde{T}(L,W)\) can be queried between simulated samples. We keep only cells with transmission above a fixed floor \(T_{\min}=0.8\). The target focusing phase is the non-paraxial hyperbolic profile
\begin{equation}
\phi_{\mathrm{tar}}(x,y)
=\left(\frac{2\pi}{\lambda}\Big[f-\sqrt{f^{2}+x^{2}+y^{2}}\Big]\right)\bmod 2\pi .
\end{equation}
We sample \(\phi_{\mathrm{tar}}(x,y)\) at the centers of the lattice cells and compare it to the library phase at each position. For each pixel \((i,j)\) we define the phase error
\begin{equation}
\Delta\phi(L,W;i,j)=\big|\tilde{\phi}(L,W)-\phi_{\mathrm{tar},ij}\big|.
\end{equation}
A unit cell is accepted if \(\Delta\phi(L,W;i,j)<\varepsilon\) with \(\varepsilon=0.07~\mathrm{rad}\) and \(\tilde{T}(L,W)\ge T_{\min}\). Among all cells that meet these two tests, we choose the one with the highest predicted transmission. We then place an SRN rectangle of height \(H\) with size \((L_{ij}^{\star},W_{ij}^{\star})\) at each lattice position, producing a layout that can be fabricated directly. Figure~\ref{fig:unitcell} summarizes the material data and the design space. Panel (a) shows refractive index \(n\) and extinction coefficient \(k\) for several SRN recipes measured by Woollam spectroscopic ellipsometry; these curves are used in the simulations. Panels (b) and (c) show, for the selected recipe at \(\lambda=660~\mathrm{nm}\), the power transmission \(T(L,W)\) and the transmission phase \(\phi(L,W)\) over the rectangular-pillar design space. Panel (d) shows the unit cell and the lateral size range \(L,W\in[150,340]~\mathrm{nm}\). The SRN deposition was tuned to achieve a relatively high refractive index in the red band, which supports a compact, subwavelength pitch and full \(2\pi\) phase coverage with high transmission. This choice is consistent with the high focusing efficiency reported in our previous work~\cite{khalilian2025high}.

% Replace the key below with your actual citation key
%\bibliographystyle{unsrt}
%\bibliography{refs}

%\author{Author One\authormark{1} and Author Two\authormark{2,*}}

%\address{\authormark{1}Peer Review, Publications Department,
%Optica Publishing Group, 2010 Massachusetts Avenue NW,
%Washington, DC 20036, USA\\
%\authormark{2}Publications Department, Optica Publishing Group,
%2010 Massachusetts Avenue NW, Washington, DC 20036, USA\\
%%\authormark{3}xyz@optica.org}

%\email{\authormark{*}xyz@optica.org}}

%Example with the corresponding author designated by an asterisk and a note indicating equal contributions by two authors.

%\author{Author One\authormark{1,3} and Author %Two\authormark{2,3,*}}

%\address{\authormark{1}Peer Review, Publications Department,
%Optica Publishing Group, 2010 Massachusetts Avenue NW, %Washington, DC 20036, USA\\
%\authormark{2}Publications Department, Optica Publishing Group, %2010 Massachusetts Avenue NW, Washington, DC 20036, USA\\
%\authormark{3}The authors contributed equally to this work.\\
%\authormark{*}xyz@optica.org}}

%\section{Examples of Article Components}
%\label{sec:examples}

\section{Fabrication}

\begin{figure}[!t]
\centering\includegraphics[width=8.7 cm]{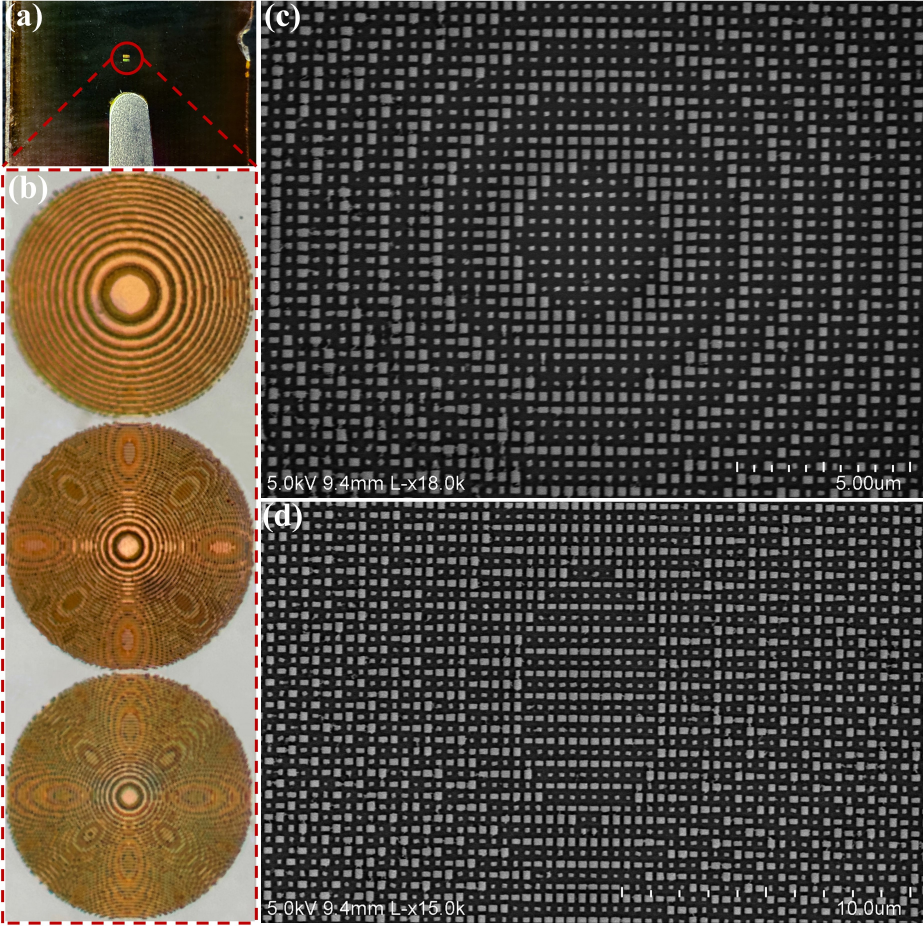}
\caption{(a) Fabricated metalens chip held by a tweezer; the array location is circled. (b) Optical micrographs of fabricated metalenses on the substrate. (c) SEM of the central lens region resolving the meta-atom lattice (scale bar 5 µm). (d) Wider-area SEM of the array (scale bar 10 µm).}
\label{fig:SEM}
\end{figure}

Fabrication proceeded on double-sided polished 4-inch fused-silica wafers of nominal 0.5\,mm thickness, which were measured individually to supply accurate inputs for ellipsometric fitting. Wafers were cleaned with IPA and briefly oxygen-plasma treated in a YES Plasma Stripper to remove organics and promote adhesion before PECVD. A 600\,nm silicon-rich nitride layer was deposited on a GSI ULTRADEP 2000 using a custom recipe, and film stress across four wafers was about 60\,MPa with no visible cracking or bowing. The wafers were diced into 1\,in\,$\times$\,1\,in coupons, then the SRN optical constants and thickness were verified with a Woollam M-2000 spectroscopic ellipsometer to match the design. Each coupon received a second short oxygen-plasma clean and a diluted ZEP520A resist coat targeting 300\,nm thickness, followed by a soft bake. A conductive topcoat (DisCharge H$_2$O) mitigated charging during electron-beam lithography, which used a dose of 240\,$\mu$C\,cm$^{-2}$ and a 500\,pA beam current. After exposure, the conductive layer was removed with DI water and the resist was developed in cold ZED-N50, rinsed in IPA, and thickness was confirmed by profilometry. A brief oxygen-plasma descum was applied, then a 60\,nm Al$_2$O$_3$ hard mask was deposited by e-beam evaporation at $\sim 4\times10^{-6}$\,Torr and 0.5\,\AA\,s$^{-1}$. Lift-off was completed in Remover PG with an extended soak and heated sonication, followed by IPA rinse. The SRN pattern was transferred using ICP-RIE in an STS APS DGRIE tool with an SF$_6$/C$_4$F$_8$ chemistry, and post-etch inspection indicated negligible Al$_2$O$_3$ residue. The fabricated sample under optical microscope and corresponding SEM images are showed in Fig.~\ref{fig:SEM}; \ref{fig:SEM}(a) shows a Fabricated metalens chip held by a tweezer; the
array location is circled, \ref{fig:SEM}(b) presents optical micrographs of the array, and \ref{fig:SEM}(c) and (d) provide SEM views that resolve the meta-atom lattice across different magnifications.

\section{Results}

\subsection{Lens characterization}
Characterization was performed on the bench shown in Fig.~2(a). For coarse alignment we used a $40\times$ Olympus objective (NA\,=\,0.95); during the focusing–spot and efficiency measurements this $40\times$ relay placed in the collection path, while the $20\times$ and $50\times$ objectives were removed because their numerical apertures are lower than the designed metalenses and would further truncate the collected cone. The device consists of a $3\times 1$ array of metalenses, each with a measured clear-aperture diameter of $60.84~\mu\mathrm{m}$. The three lenses exhibit numerical apertures of 0.54, 0.92, and 0.97 and produce focal-spot FWHM values of $0.92~\mu\mathrm{m}$, $0.84~\mu\mathrm{m}$, and $0.83~\mu\mathrm{m}$, respectively, as summarized in Fig.~2(f). The corresponding focusing efficiencies are $65.9\%$, $20.5\%$, and $7.0\%$, where “focusing efficiency” denotes the ratio of background-subtracted power integrated within a circular aperture of diameter $3\times\mathrm{FWHM}$, centered on the focal peak, to the total background-subtracted power transmitted through the lens and collected on the same plane. The efficiency reported for the highest-NA element is partially limited by the NA\,0.95 relay: when the metalens NA approaches or exceeds the collection NA, the captured solid angle is under-filled and the measured value is a lower bound relative to the incident power.

\begin{figure}[!t]
\centering\includegraphics[width=8.7 cm]{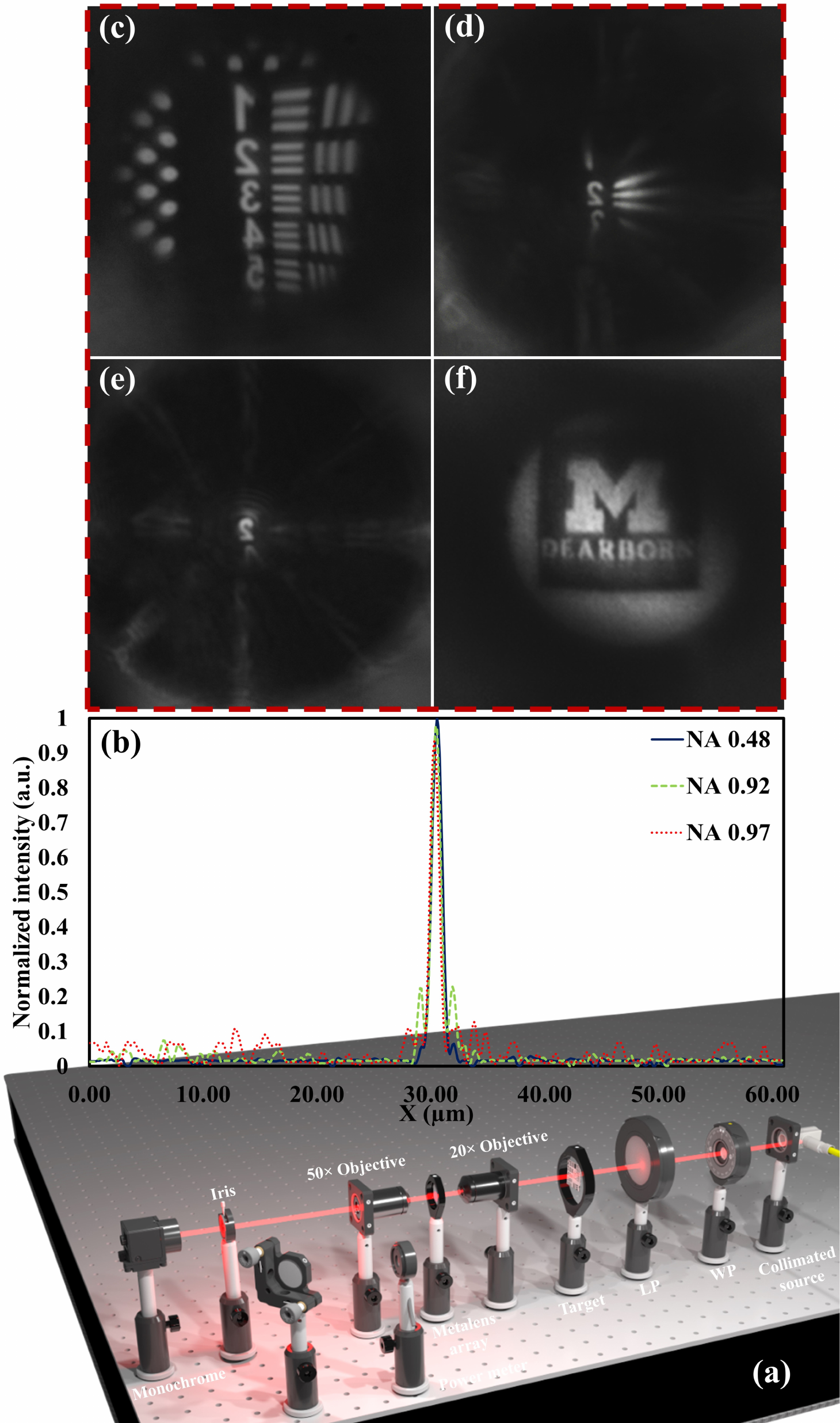}
\caption{(a) Overview of the imaging setup. (b) Normalized intensity profiles at focus for NA 0.54, 0.92, and 0.97, showing a narrower main lobe with increasing NA. (c–e) Imaging of a USAF 1951 high-resolution target through the metalens array at different numerical apertures. Higher NA yields crisper group elements and resolves the “2” feature more distinctly. (f) Imaging of a test logo for qualitative verification of NA 0.54 lens.}
\label{fig:target imaging}
\end{figure}

Focal distances were inferred from the measured numerical apertures for the common diameter of $60.84~\mu\mathrm{m}$, yielding $f=47.41~\mu\mathrm{m}$ for NA 0.54, $f=12.96~\mu\mathrm{m}$ for NA 0.92, and $f=7.62~\mu\mathrm{m}$ for NA 0.97. The university-logo demonstration in Fig.~2(e) uses a lens operated at NA $\approx 0.52$, which corresponds to $f=49.97~\mu\mathrm{m}$ for the same diameter. These focal distances were used to position axial monitors and to set image planes during measurement. Imaging performance on test targets was then evaluated at the appropriate conjugates for each focal length; representative images for the high-, intermediate-, and low-NA elements appear in Fig.~\ref{fig:target imaging}(b), (e), and (d), respectively, with Fig.~\ref{fig:target imaging}(e) showing the university logo formed by the NA $\approx 0.52$ lens.

\begin{figure}[!t]
\centering\includegraphics[width=8.7 cm]{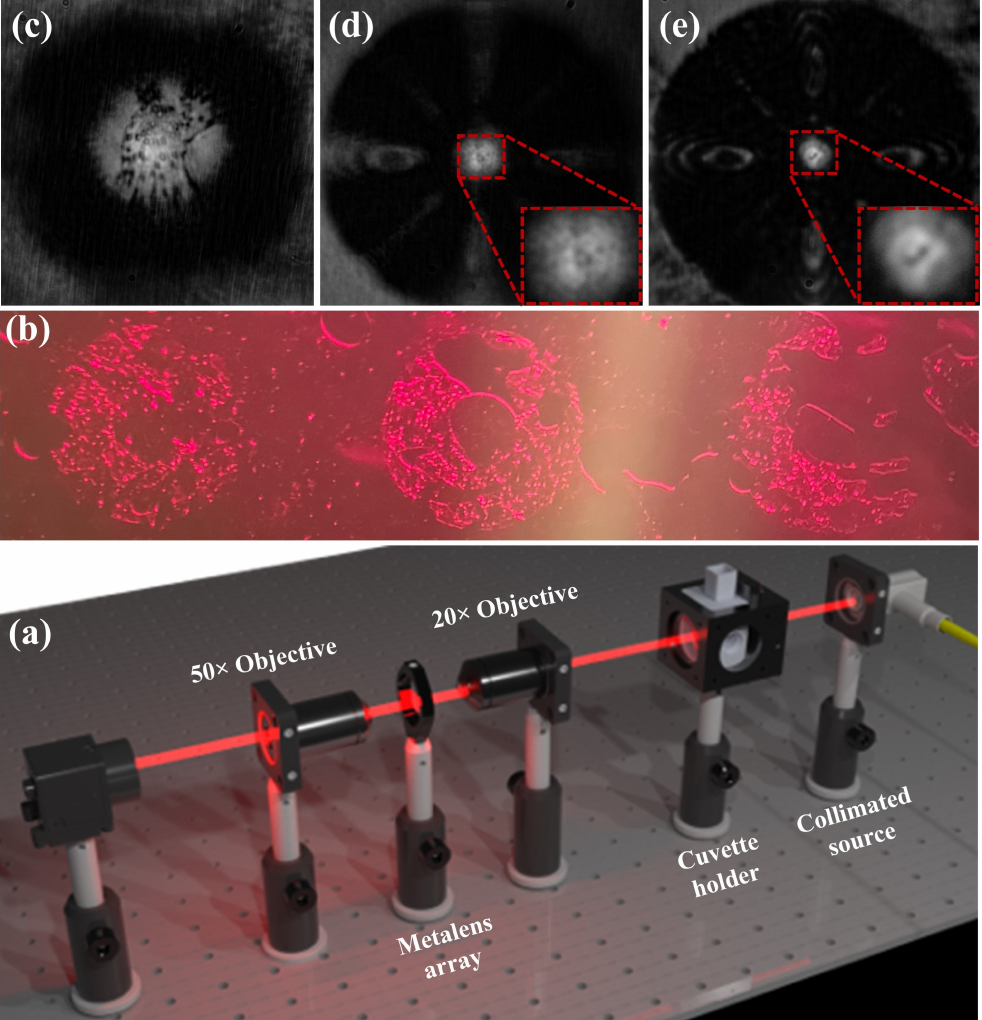}
\caption{(a) Bench setup with the metalens array between relay objectives. (b) Annular cell monolayers on coverslips prior to fluorescence imaging. (c) AF647 ZO-1 fluorescence of the annular ring acquired with the NA 0.54 metalens. (d) Single-cell AF647 ZO-1 fluorescence acquired with the NA 0.92 metalens. (e) Single-cell AF647 ZO-1 fluorescence acquired with the NA 0.97 metalens, showing the highest junctional edge clarity.}
\label{fig:cell imaging}
\end{figure}

\subsection{Cell microscopy}

We performed fluorescence imaging of annular cell monolayers using 642\,nm excitation and a 650\,nm long-pass emission filter, recording only the ZO-1 immunolabel in the far-red via Alexa Fluor 647. In this study, we focus on cell–cell junctions; ZO-1 tagged with AF647 marks tight junctions along the interfaces between adjacent cells. The bench arrangement is shown in Fig.~\ref{fig:cell imaging}(a). Figure~\ref{fig:cell imaging}(b) shows the annular monolayers on coverslips prior to AF647 acquisition; rings were produced by microcontact printing of adhesive, ring-shaped micropatterns on Au-coated glass (PDMS stamp from an SU-8 master; ODT/EG3 surface chemistry; fibronectin coating) to confine cells to an annulus and promote uniform monolayer formation \cite{fan2018cell}. The annular geometry provides an axially uniform in-focus layer, suppresses central background, and offers two sharp edges that serve as on-sample resolution gauges. Figures~\ref{fig:cell imaging}(c)–\ref{fig:cell imaging}(e) present AF647 fluorescence from the ZO-1 channel only, recorded with three metalenses on the same chip with numerical apertures 0.54, 0.92, and 0.97. Figure~\ref{fig:cell imaging}(c) shows the full annulus with the 0.54 lens, where the seeded ring appears bright against a dark background and both inner and outer edges are visible. Figures~\ref{fig:cell imaging}(d) and \ref{fig:cell imaging}(e) show single-cell views captured with the 0.92 and 0.97 lenses; junctional contours sharpen with increasing numerical aperture, consistent with Rayleigh scaling \(0.61\,\lambda/\mathrm{NA}\) at \(\lambda\approx660~\mathrm{nm}\). For these imaging experiments the metalens output was relayed to the camera using $20\times$ or $50\times$ objectives, so the effective system NA is the minimum of the metalens NA and the relay objective NA. As a result, the reported image resolutions are conservative; a direct lens-on-sensor configuration would permit the intrinsic metalens NA to dictate resolution but is impractical here due to the micrometer-scale apertures in the array and fabrication constraints. The metasurface platform uses silicon-rich nitride with low loss near 660\,nm and a visible refractive index sufficient for full \(2\pi\) phase control at a single etched height \(H=600~\mathrm{nm}\). The PECVD plus single dry-etch flow is low temperature, CMOS compatible, and wafer scalable, enabling multi-NA arrays on one substrate for compact bioimaging modules.

\begin{backmatter}

\bmsection{Acknowledgment} The authors thank the Lurie Nanofabrication Facility (LNF) at the University of Michigan for support with device fabrication.

\bmsection{Disclosures} The authors declare no conflicts of interest.

\bmsection{Data availability} Data underlying the results presented in this paper are not publicly available at this time but may be obtained from the authors upon reasonable request.

See Supplement 1 for supporting content

\end{backmatter}

% Bibliography
\bibliography{sample}

% Full bibliography added automatically for Optics Letters submissions; the following line will simply be ignored if submitting to other journals.
% Note that this extra page will not count against page length
\bibliographyfullrefs{sample}

%Manual citation list
%\begin{thebibliography}{1}
%\bibitem{Zhang:14}
%Y.~Zhang, S.~Qiao, L.~Sun, Q.~W. Shi, W.~Huang, %L.~Li, and Z.~Yang,
 % \enquote{Photoinduced active terahertz metamaterials with nanostructured
  %vanadium dioxide film deposited by sol-gel method,} Opt. Express \textbf{22},
  %11070--11078 (2014).
%\end{thebibliography}

\end{document}